\documentclass[aps,prl,groupedaddress,fleqn,twocolumn]{revtex4}
\pdfoutput=1
\usepackage{graphicx}
\usepackage{amsmath}
\usepackage{gensymb}
\usepackage{physics}

\begin{document}

\title{Quantum dissipation in a scalar field theory with gapped momentum states}
\author{K. Trachenko}
\address{School of Physics and Astronomy, Queen Mary University of London, Mile End Road, London, E1 4NS, UK. Correspondence to k.trachenko@qmul.ac.uk}

\begin{abstract}
Understanding quantum dissipation is important from both theoretical perspective and applications. Here, we show how to describe dissipation in a scalar field theory. We treat dissipation non-perturbatively, represent it by a bilinear term in the Lagrangian and quantize the theory. We find that dissipation promotes a gap in momentum space and reduces the particle energy. As a result, particle mass becomes dressed by dissipation due to self-interaction. The underlying mechanism is similar to that governing the propagation of transverse collective modes in liquids. We discuss the interplay between the dissipative and mass terms, the associated different regimes of field dynamics and the emergence of ultraviolet and infrared cutoffs due to dissipation.
\end{abstract}


\maketitle

Theoretical description of dissipation in quantum systems is an interesting and challenging problem of fundamental importance related to the foundations of quantum theory itself (see, e.g., \cite{bender,mohsen}). Quantum dissipation has seen renewed recent interest in areas related to non-equilibrium and irreversible physics, decoherence effects and complex systems. Starting from early work (see, e.g., \cite{feynman,leggett}), a common approach to treat dissipation is to introduce a central dissipative system of interest, its environment modelled as, for example, a bath of harmonic oscillators and an interaction between the system and its environment enabling energy exchange (see, e.g., \cite{kamenev,weiss} for review). In this picture, dissipative effects can be discussed by solving models using approximations such as linearity of the system and its couplings.

Here, we propose a conceptually different description of dissipation based on recent insights of wave propagation in liquids and supercritical fluids \cite{ropp,prl,pre}. No dissipation takes place when a plane wave propagates in a crystal where the wave is an eigenstate. However, a plane wave dissipates in systems with structural and dynamical disorder such as liquids. An important result is the emergence of the gap in $k$- or momentum space in the transverse wave spectrum, with the accompanying decrease of the wave energy due to dissipation. We propose that this is a physically relevant mechanism to introduce and discuss dissipation in quantum field theory in a general way because canonical quantization of fields involves expanding field operators in terms of plane waves.

In this paper, we show how dissipation can be introduced in a scalar field theory. Representing dissipation due to field self-interaction by a bilinear term as in liquids, we perform canonical quantization of scalar fields. We find that particle energy is reduced by dissipation which promotes the gapped momentum state (GMS). Particle mass becomes dressed by dissipation as a result. The underlying mechanism is similar to that governing transverse modes in liquids. We discuss the interplay between the dissipative and mass terms, the associated different regimes of field dynamics as well as ultraviolet and infrared cutoffs due to dissipation.

We start with recalling how liquid transverse modes develop a GMS and how this effect can be represented by a Lagrangian. We note first-principles description of liquids is exponentially complex and is not tractable because it involves a large number of coupled non-linear oscillators \cite{ropp}. At the same time, liquids have no simplifying small parameters as in gases and solids \cite{landau}. However, progress in understanding liquid modes can be made by using non-perturbative approach to liquids pioneered by Maxwell and developed later by Frenkel. This programme involves Maxwell interpolation:

\begin{equation}
\frac{ds}{dt}=\frac{P}{\eta}+\frac{1}{G}\frac{dP}{dt}
\label{a1}
\end{equation}
\noindent where $s$ is shear strain, $\eta$ is viscosity, $G$ is shear modulus and $P$ is shear stress.

(\ref{a1}) reflects Maxwell's proposal \cite{maxwell} that shear response in a liquid is the sum of viscous and elastic responses given by the first and second right-hand side terms. Notably, the dissipative term containing viscosity is not introduced as a small perturbation: both elastic and viscous deformations are treated in (\ref{a1}) on equal footing.

Frenkel proposed \cite{frenkel} to represent the Maxwell interpolation (\ref{a1}) by introducing the operator $A=1+\tau\frac{d}{dt}$ and write Eq. (\ref{a1}) as $\frac{ds}{dt}=\frac{1}{\eta}AP$. Here, $\tau$ is Maxwell relaxation time $\frac{\eta}{G}$. Frenkel's theory has identified $\tau$ with the average time between consecutive molecular jumps in the liquid \cite{frenkel}. This has become an accepted view \cite{dyre}. Frenkel's idea was to generalize $\eta$ to account for liquid's short-time elasticity as $\frac{1}{\eta}\rightarrow\frac{1}{\eta}\left(1+\tau\frac{d}{dt}\right)$ and use this $\eta$ in the Navier-Stokes equation $\nabla^2{\bf v}=\frac{1}{\eta}\rho\frac{d{\bf v}}{dt}$, where ${\bf v}$ is velocity, $\rho$ is density and $\frac{d}{dt}=\frac{\partial}{\partial t}+{\bf v\nabla}$. We have carried this idea forward \cite{ropp} and, considering small ${\bf v}$, wrote:

\begin{equation}
c^2\frac{\partial^2v}{\partial x^2}=\frac{\partial^2v}{\partial t^2}+\frac{1}{\tau}\frac{\partial v}{\partial t}
\label{gener3}
\end{equation}

\noindent where $v$ is the velocity component perpendicular to $x$, $\eta=G\tau=\rho c^2\tau$ and $c$ is the shear wave velocity.

Eq. (\ref{gener3}) can also be obtained by starting with the solid-like equation for the non-decaying wave and, using Maxwell interpolation (\ref{a1}), generalizing the shear modulus to include the viscous response \cite{pre}.

In contrast to the Navier-Stokes equation, (\ref{gener3}) contains the second time derivative and hence gives propagating waves. We solved Eq. (\ref{gener3}) in Ref. \cite{ropp}: seeking the solution as $v=v_0\exp\left(i(kx-\omega t)\right)$ gives $\omega^2+\omega\frac{i}{\tau}-c^2k^2=0$ and

\begin{equation}
\begin{aligned}
v\propto\exp\left(-\frac{t}{2\tau}\right)\exp(i\omega t)\\
\omega=\sqrt{c^2k^2-\frac{1}{4\tau^2}}
\end{aligned}
\label{omega}
\end{equation}

$\tau\rightarrow\infty$ in (\ref{omega}) and the absence of dissipation in (\ref{gener3}) correspond to an infinite range of propagation of the plane shear wave as in the crystal. A finite $\tau$ implies dissipation of the wave in a sense that it acquires a {\it finite} propagation range. Indeed, the dissipation takes place over time approximately equal to $\tau$ according to (\ref{omega}). This corresponds to the propagation range of the shear wave being finite and equal to $c\tau$ (this can be inferred directly from the Frenkel's discussion: if $\tau$ is the time during which shear stress can exist in a liquid, $c\tau$ is the distance over which a shear wave propagates.) $\tau$ sets the physical time scale during which we consider the dissipation process: if an observation of an injected shear wave starts at $t=0$, time $t\approx\tau$ is the end of the process because over this time the wave amplitude and energy appreciably reduce.

An important property is the emergence of the gap in $k$-space or GMS: in order for $\omega$ in (\ref{omega}) to be real, $k>k_g$ should hold, where $k_g=\frac{1}{2c\tau}$ increases with temperature because $\tau$ decreases. Recently \cite{prl}, detailed evidence for GMS was presented on the basis of molecular dynamics simulations. Figure 1 illustrates these findings.

\begin{figure}
\begin{center}
{\scalebox{0.35}{\includegraphics{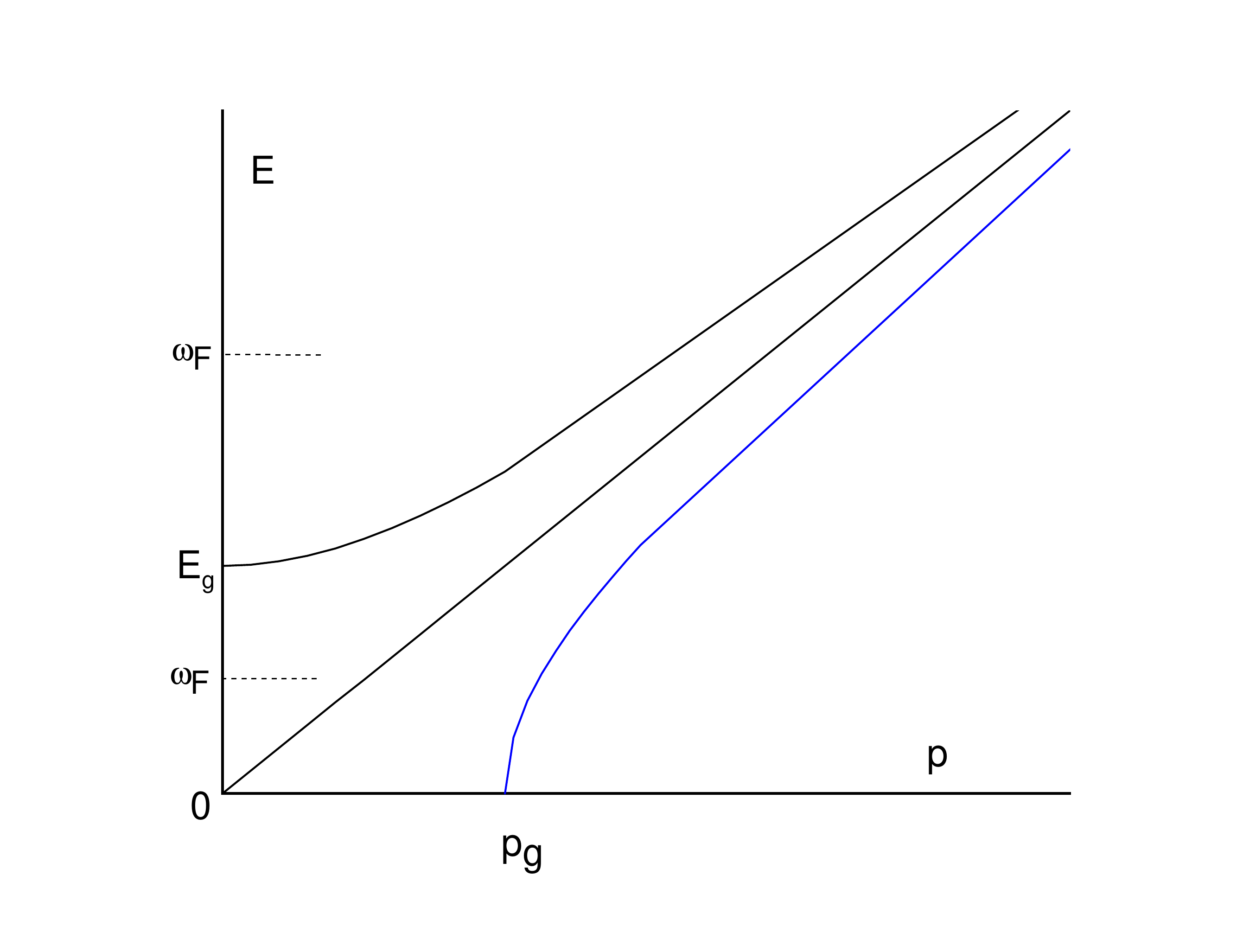}}}
\end{center}
\caption{Possible dispersion relations and dependencies of energy $E$ on momentum $p$. Top curve shows the dispersion relation for a massive particle. Middle curve shows gapless dispersion relation for a massless particle (photon) or a phonon in solids. Bottom curve shows the dispersion relation (\ref{omega}) with the gap in $k$-space, illustrating the results of Ref. \cite{prl}. Two values of $\omega_{\rm F}=\frac{1}{\tau}$ in relation to the energy gap $E_g$ are shown.
}
\label{3}
\end{figure}

The GMS is interesting. Indeed, the two commonly discussed types of dispersion relations are either gapless as for photons and phonons, $E=p$ ($c=1$), or have the energy gap for massive particles, $E=\sqrt{p^2+m^2}$, where the gap is along the Y-axis. On the other hand, (\ref{omega}) implies that the gap is in {\it momentum} space and along the X-axis, similar to the hypothesized tachyon particles with imaginary mass \cite{tachyons}.

It has been realized that in addition to liquids, GMS emerge in a surprising variety of areas \cite{review}, including strongly-coupled plasma, electromagnetic waves, non-linear Sine-Gordon model, relativistic hydrodynamics and holographic models.

An important question from field-theoretical perspective is what Lagrangian gives Eq. (\ref{gener3}) and the associated GMS? The challenge is to represent the viscous term $\propto\frac{1}{\tau}$ in (\ref{gener3}) in the Lagrangian. The viscous energy can be written as the work $W$ done to move the liquid. If $s$ is the strain, $W\propto Fs$, where $F$ is the viscous force $F\propto\eta\frac{ds}{dt}$. Hence, the dissipative term in the Lagrangian should contain the term $s\frac{ds}{dt}$. This can be represented by a scalar field $\phi$, giving the term $L\propto\phi\frac{\partial\phi}{\partial t}$. However, the term $\phi\frac{d\phi}{dt}$ disappears from the Euler-Lagrange equation $\frac{\partial L}{\partial\phi}=\frac{\partial}{\partial t}\frac{\partial L}{\partial\frac{\partial\phi}{\partial t}}+\frac{\partial}{\partial x}\frac{\partial L}{\partial\frac{\partial\phi}{\partial x}}$ because $\frac{\partial L}{\partial\phi}=\frac{\partial}{\partial t}\frac{\partial L}{\partial\frac{\partial\phi}{\partial t}}=\frac{\partial\phi}{\partial t}$. Another way to see this is note that the viscous term $\phi\frac{d\phi}{dt}\propto\frac{d}{dt}\phi^2$.

To circumvent this problem, we proposed to operate in terms of {\it two} fields $\phi_1$ and $\phi_2$ \cite{pre} and constructed the dissipative term as a combination of $\phi\frac{d\phi}{dt}$, namely as $L_d=\frac{1}{2\tau}\left(\phi_1\frac{\partial\phi_2}{\partial t}-\phi_2\frac{\partial\phi_1}{\partial t}\right)$. We note that a two-coordinate description of a localised damped harmonic oscillator was discussed earlier \cite{bateman,dekker}. Increasing the number of degrees of freedom was also involved in describing dissipation with gapless dispersion relations and in the hydrodynamic regime only \cite{grozdanov}.

The Lagrangian becomes:

\begin{equation}
\begin{split}
&L=\frac{\partial\phi_1}{\partial t}\frac{\partial\phi_2}{\partial t}-\frac{\partial\phi_1}{\partial x}\frac{\partial\phi_2}{\partial x}+\\
&\frac{1}{2\tau}\left(\phi_1\frac{\partial\phi_2}{\partial t}-\phi_2\frac{\partial\phi_1}{\partial t}\right)-m^2\phi_1\phi_2
\label{l1}
\end{split}
\end{equation}
\noindent where we added the mass term, assumed $c=1$ and, for simplicity, considered one-dimensional case.

We note that (\ref{l1}) without the mass term follows from the two-field Lagrangian

\begin{equation}
\begin{split}
&L=\frac{1}{2}\left(\left(\frac{\partial\psi_1}{\partial t}\right)^2-\left(\frac{\partial\psi_1}{\partial x}\right)^2+\left(\frac{\partial\psi_2}{\partial t}\right)^2-\left(\frac{\partial\psi_2}{\partial x}\right)^2\right)\\
&+\frac{i}{2\tau}\left(\psi_2\frac{\partial\psi_1}{\partial t}-\psi_1\frac{\partial\psi_2}{\partial t}\right)
\label{l11}
\end{split}
\end{equation}

\noindent using the transformation employed in the complex field theory: $\phi_1=\frac{1}{\sqrt{2}}(\psi_1+i\psi_2)$ and $\phi_2=\frac{1}{\sqrt{2}}(\psi_1-i\psi_2)$. The advantage of using (\ref{l1}) in terms of $\phi_1$ and $\phi_2$ is that the equations of motion for $\phi_1$ and $\phi_2$ decouple as we see below. This is not an issue, however: one can use (\ref{l11}) to obtain the system of coupled equations for $\psi_1$ and $\psi_2$ and decouple them using the same transformation between $\phi$ and $\psi$, resulting in the same equations for $\phi$ as those following from (\ref{l1}). Note that the imaginary term in (\ref{l11}) may be related to dissipation \cite{bender,sudarshan,sud1,qft}, however the Hamiltonian corresponding to (\ref{l11}) does not explicitly contain an imaginary term: $H=\frac{1}{2}\left(\left(\frac{\partial\psi_1}{\partial t}\right)^2+
\left(\frac{\partial\psi_1}{\partial x}\right)^2+\left(\frac{\partial\psi_2}{\partial t}\right)^2+\left(\frac{\partial\psi_2}{\partial x}\right)^2\right)$, where terms with $\tau$ cancel out and where the real parts of $\psi_1$ and $\psi_2$ are implied.

$\tau\rightarrow\infty$ corresponds to no particle jumps and $L_d=0$, in which case $L$ in (\ref{l1}) takes the form of the complex scalar field theory (this corresponds to the short-time regime $t\ll\tau$ in the solution of (\ref{l1}) as shown below). The dissipative term $\propto\frac{1}{\tau}$ in (\ref{l1}) and (\ref{l11}) can be viewed as a coupling term between two sectors describing $\phi_1$ and $\phi_2$ ($\psi_1$ and $\psi_2$). As discussed below, this coupling results in the flow of energy from one field to another. The coupling breaks time reversal symmetry of Lagrangian $t\rightarrow -t$, however the Lagrangian is invariant under the combination of time reversal symmetry and internal symmetry $\phi_1\rightarrow\phi_2$ ($\psi_1\rightarrow\psi_2$). Combinations of spacetime and internal symmetries in condensed matter systems are discussed more generally, e.g., in Ref. \cite{symbrek}.

We now consider the Hamiltonian of the dissipative system and its quantization. Applying the Euler-Lagrange equation to (\ref{l1}) gives two decoupled equations $\frac{\partial^2\phi_1}{\partial x^2}=\frac{\partial^2\phi_1}{\partial t^2}+\frac{1}{\tau}\frac{\partial\phi_1}{\partial t}+m^2\phi_1$ and $\frac{\partial^2\phi_2}{\partial x^2}=\frac{\partial^2\phi_2}{\partial t^2}-\frac{1}{\tau}\frac{\partial\phi_2}{\partial t}+m^2\phi_2$, where the equation for $\phi_1$ is the same as (\ref{gener3}) when $m=0$. A solution for $\phi_1$ and $\phi_2$ can be written as:

\begin{equation}
\begin{aligned}
\phi_1=\phi_0e^{i(E_pt-px)}e^{-\frac{t}{\tau}}\\
\phi_2=\phi_0e^{-i(E_pt-px)}e^{\frac{t}{\tau}}
\end{aligned}
\label{2phidag}
\end{equation}

\noindent with

\begin{equation}
E_p=\sqrt{p^2+m^2-\frac{1}{\tau^2}}
\label{energydis}
\end{equation}

\noindent where, for simplicity, we absorbed the factor of $2$ in $\tau$.

$E_p$ in (\ref{energydis}) is the same as in (\ref{omega}) for $m=0$. Importantly and similarly to (\ref{omega}), $E_p$ is reduced from $E_p=\sqrt{p^2+m^2}$ due to the dissipative term $\propto\frac{1}{\tau}$ and has a gap in momentum space as discussed below in detail.

$\phi_1$ and $\phi_2$ in (\ref{2phidag}) decrease and increase with time, respectively. This stems from choosing a positive sign of the dissipative term $L_d=\frac{1}{2\tau}\left(\phi_1\frac{\partial\phi_2}{\partial t}-\phi_2\frac{\partial\phi_1}{\partial t}\right)$ in (\ref{l1}) (a negative sign results in $\phi_1\propto e^{\frac{t}{\tau}}$ and $\phi_2\propto e^{-\frac{t}{\tau}}$ in (\ref{2phidag})). $\phi_1$ and $\phi_2$ in (\ref{2phidag}) can be viewed as energy exchange between waves $\phi_1$ and $\phi_2$: $\phi_1$ and $\phi_2$ appreciably decrease and grow over time $\tau$, respectively. This process is not dissimilar from phonon scattering in crystals due to defects or anharmonicity where a plane-wave phonon ($\phi_1$) decays into other phonons (represented by $\phi_2$) and acquires a finite lifetime $\tau$ as a result.

The momenta from (\ref{l1}) are (recalling that $2\tau\rightarrow\tau$):

\begin{equation}
\begin{aligned}
\pi_1=\frac{\partial\phi_2}{\partial t}-\frac{\phi_2}{\tau}\\
\pi_2=\frac{\partial\phi_1}{\partial t}+\frac{\phi_1}{\tau}
\end{aligned}
\label{momenta}
\end{equation}

\noindent and the Hamiltonian density is

\begin{equation}
{\mathcal H}=\frac{\partial\phi_1}{\partial t}\frac{\partial\phi_2}{\partial t}+\frac{\partial\phi_1}{\partial x}\frac{\partial\phi_2}{\partial x}+m^2\phi_1\phi_2
\label{h11}
\end{equation}

\noindent where terms $\propto\frac{1}{\tau}$ cancel out.

We now proceed to quantization.  (\ref{2phidag}) are solutions of the dissipative Lagrangian (\ref{l1}). Setting $\tau\rightarrow\infty$ in (\ref{l1}) and considering the short-time regime $t\ll\tau$ in (\ref{2phidag}) corresponds to the absence of dissipation and to the complex scalar field theory, which is quantized using operators acting on particles and antiparticles \cite{qft}. Therefore, it is convenient to quantize the full Hamiltonian

\begin{equation}
H=\int d^3x\left(\frac{\partial\phi_1}{\partial t}\frac{\partial\phi_2}{\partial t}+\frac{\partial\phi_1}{\partial x}\frac{\partial\phi_2}{\partial x}+m^2\phi_1\phi_2\right)
\label{h1}
\end{equation}

\noindent by using the canonical quantization of the complex field theory and augment it with factors $e^{\pm\frac{t}{\tau}}$ in (\ref{2phidag}):

\begin{equation}
\begin{aligned}
\phi_1=\int\frac{d^3p}{(2\pi)^\frac{3}{2}}\frac{1}{(2E_p)^\frac{1}{2}}\left(a_pe^{-ipx}+b_p^\dagger e^{ipx}\right)e^{-\frac{t}{\tau}}\\
\phi_2=\int\frac{d^3p}{(2\pi)^\frac{3}{2}}\frac{1}{(2E_p)^\frac{1}{2}}\left(a_p^\dagger e^{ipx}+b_p e^{-ipx}\right)e^{\frac{t}{\tau}}
\end{aligned}
\label{modexp}
\end{equation}

In (\ref{modexp}), operators $a$ and $b$ act on particles and antiparticles as usual and $p$ and $x$ are four-vectors. The normalizing factors follow from the mass-shell condition as usual, albeit involving $E_p$ in (\ref{energydis}). In the short-time regime $t\ll\tau$ where dissipation can be neglected, (\ref{modexp}) is the mode expansion used to quantize the complex scalar field theory \cite{qft}.

As in (\ref{omega}), the time scale over which we consider and describe the dissipation process in both (\ref{2phidag}) and (\ref{modexp}) is $\tau$ because the phonon with the $k$-gap (\ref{omega}) dissipates after time comparable to $\tau$.

A persisting open problem in using quantum mechanics to describe dissipation involved two dual Bateman oscillators where commutation relations did not hold \cite{dekker}. On the other hand, the quantization based on the mode expansion proposed here satisfies the required commutation relations for the field operators. The commutators $[\phi_1,\phi_1]$, $[\phi_2,\phi_2]$, $[\phi_2,\pi_1]$, $[\phi_1,\pi_2]$, $[\pi_1,\pi_1]$ and $[\pi_2,\pi_2]$ do not involve $[a_p,a_p^\dagger]$ or $[b_p,b_p^\dagger]$ and are zero. $[\phi_1,\phi_2]$ is zero because the exponential factors $e^{-\frac{t}{\tau}}$ and $e^{\frac{t}{\tau}}$ in (\ref{modexp}) cancel out, resulting in $[\phi_1,\phi_2]=0$ for the same reason as in the standard complex field theory. The commutator $[\phi_1(x),\pi_1(y)]$ is $[\phi_1(x),\frac{\partial\phi_2(y)}{\partial t}-\frac{\phi_2(y)}{\tau}]$ from (\ref{momenta}), or $[\phi_1(x),\frac{\partial\phi_2(y)}{\partial t}]$ because $\phi_1$ and $\phi_2$ commute. Considering equal times and using (\ref{modexp}), $[\phi_1(x),\pi_1(y)]=\int\frac{d^3p}{(2\pi)^3}\frac{1}{2E_p}[a_pe^{-i{\bf px}}+b_p^\dagger e^{i{\bf px}},a_p^\dagger e^{i{\bf py}}\left(iE_p+\frac{1}{\tau}\right)+b_pe^{-i\bf{py}}\left(-iE_p+\frac{1}{\tau}\right)]$, giving the commutator in the integrand as $e^{-i\bf{p}(\bf{x}-\bf{y})}\left(iE_p+\frac{1}{\tau}\right)+e^{i\bf{p}(\bf{x}-\bf{y})}\left(iE_p-\frac{1}{\tau}\right)$. Swapping the sign of $\bf{p}$ in the first term and integrating gives $[\phi_1(x),\pi_1(y)]=i\delta(\bf{x}-\bf{y})$ as required because the terms with $\frac{1}{\tau}$ cancel. The same cancellation mechanism gives $[\phi_2(x),\pi_2(y)]=i\delta(\bf{x}-\bf{y})$. Finally, $[\pi_1(x),\pi_2(y)]=[\frac{\partial\phi_2(x)}{\partial t}-\frac{\phi_2(x)}{\tau},\frac{\partial\phi_1(y)}{\partial t}+\frac{\phi_1(y)}{\tau}]=[\frac{\partial\phi_2(x)}{\partial t},\frac{\partial\phi_1(y)}{\partial t}]+\frac{1}{\tau}\left([\frac{\partial\phi_2(x)}{\partial t},\phi_1(y)]+[\frac{\partial\phi_1(y)}{\partial t},\phi_2(x)]\right)$. The first term contains $e^{i\bf{p}(\bf{x}-\bf{y})}\left(iE_p+\frac{1}{\tau}\right)^2-e^{-i\bf{p}(\bf{x}-\bf{y})}\left(iE_p-\frac{1}{\tau}\right)^2$, contributing $\frac{2i}{\tau}\delta(\bf{x}-\bf{y})$ to $[\pi_1(x),\pi_2(y)]$. The second term contributes $-\frac{2i}{\tau}\delta(\bf{x}-\bf{y})$ to $[\pi_1(x),\pi_2(y)]$, using the earlier results for $[\phi_1(y),\frac{\partial\phi_2(x)}{\partial t}]$ and $[\phi_2(x),\frac{\partial\phi_1(y)}{\partial t}]$. The two terms cancel, giving $[\pi_1(x),\pi_2(y)]=0$ as required.

Using (\ref{modexp}) in (\ref{h1}) gives

\begin{equation}
\begin{aligned}
H=\int\frac{d^3p}{2E_p}\left(E_p^2+p^2+m^2-\frac{1}{\tau^2}-\frac{2iE_p}{\tau}\right)a_pa_p^\dagger+\\
\left(E_p^2+p^2+m^2-\frac{1}{\tau^2}+\frac{2iE_p}{\tau}\right)b_pb_p^\dagger+\\
\left(-E_p^2+p^2+m^2-\frac{1}{\tau^2}\right)a_pb_{-p}e^{-2E_pit}+\\
\left(-E_p^2+p^2+m^2-\frac{1}{\tau^2}\right)a_{-p}^\dagger b_p^\dagger e^{2E_pit}
\end{aligned}
\label{quant1}
\end{equation}

Using (\ref{energydis}), (\ref{quant1}) simplifies, after normal ordering, to

\begin{equation}
H=\int d^3p E_p\left(a_p^\dagger a_p+b_p^\dagger b_p\right)-i\omega_{\rm F}\int d^3p(a_p^\dagger a_p-b_p^\dagger b_p)
\label{q3}
\end{equation}

\noindent with the energy spectrum (\ref{energydis})

\begin{equation}
E_p=\sqrt{p^2+m^2-\omega_{\rm F}^2}
\label{en-omega}
\end{equation}

\noindent where $\omega_{\rm F}=\frac{1}{\tau}$ is the Frenkel hopping frequency.

Eqs. (\ref{q3})-(\ref{en-omega}) represent a canonically quantized theory with dissipation. The first term in (\ref{q3}) describes the excitations of particles and antiparticles as in the complex scalar field theory, albeit with energies reduced by the dissipation according to (\ref{en-omega}) due to the presence of the field hopping process with frequency $\omega_{\rm F}$ (see below for a more detailed discussion). If the number of particles $N_p=\int d^3pa_p^\dagger a_p$ and the number of antiparticles $N_a=\int d^3pb_p^\dagger b_p$ are equal, the second term in (\ref{q3}) is zero. In this case, (\ref{q3}) is real, corresponding to the stationary state of the system. $N_a\ne N_p$ in (\ref{q3}) corresponds to non-zero imaginary term, which is related to particle decay and dissipation \cite{qft,sudarshan,sud1}. We will discuss the relationship between the dissipation and the anharmonic interaction potential below.

We note that similarly to (\ref{2phidag}), there is a choice of assigning factors $e^{-\frac{t}{\tau}}$ and $e^{\frac{t}{\tau}}$ to $\phi_1$ and $\phi_2$ in (\ref{modexp}). If $\phi_1\propto e^{\frac{t}{\tau}}$ and $\phi_2\propto e^{-\frac{t}{\tau}}$, the imaginary term in (\ref{q3}) changes sign.

Time-dependent properties of this model will be discussed elsewhere. Here, we note that the dissipation is not related to a change of the total energy of the system (a conserved property); rather, it is related to the dissipation of harmonic excitations in the strongly-anharmonic potential as is the case of dissipation of plane waves in structurally and dynamically disordered liquids where a plane wave acquires a finite lifetime and propagation range as discussed earlier.

\begin{figure}
\begin{center}
{\scalebox{0.27}{\includegraphics{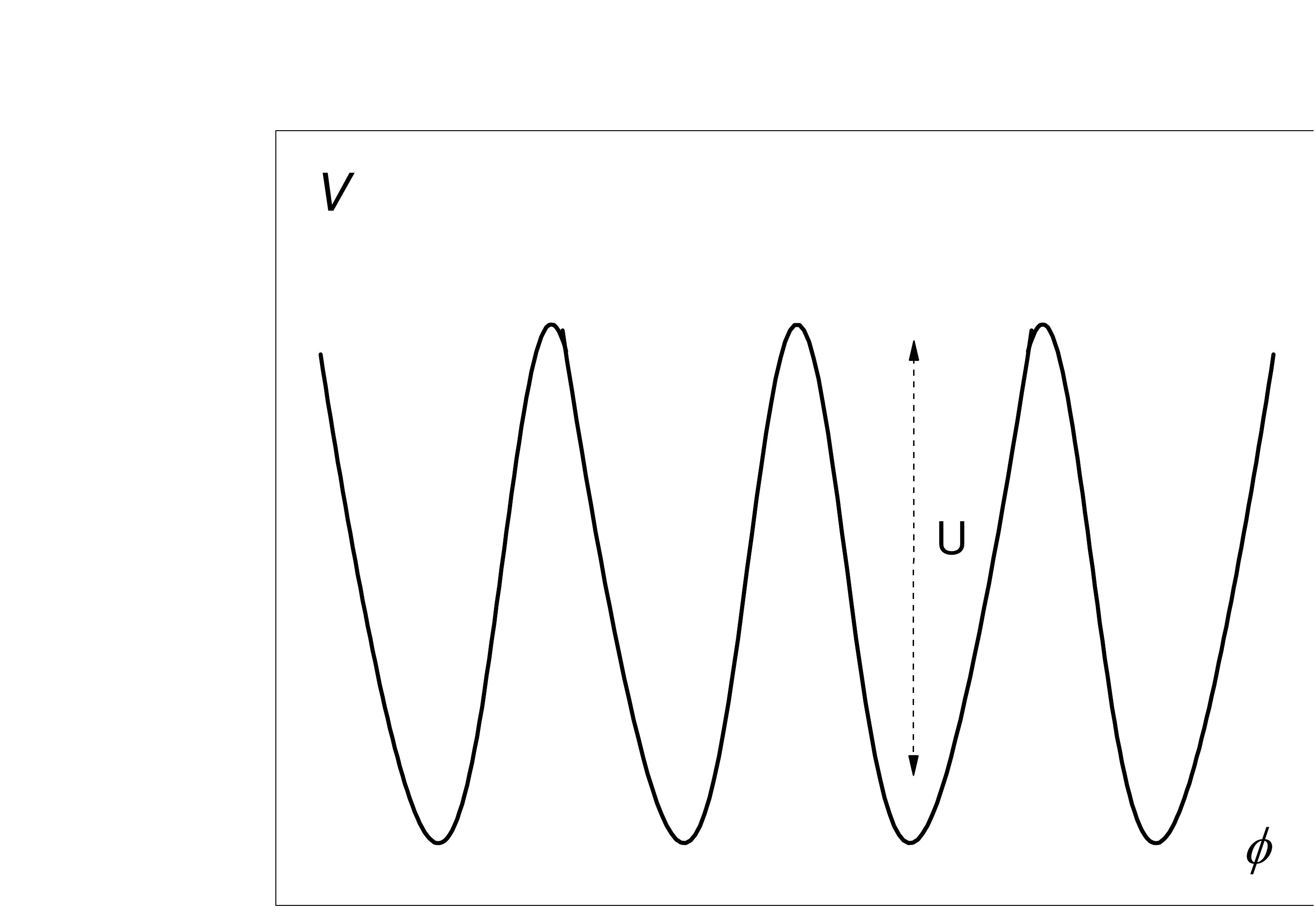}}}
\end{center}
\caption{Schematic illustration of an interaction potential.
}
\label{poten}
\end{figure}

We pause for the moment and comment on the canonical quantization performed in (\ref{modexp})-(\ref{en-omega}). Canonical quantization is believed to be possible only for Lagrangians quadratic in fields and its derivatives, with the result that the system is the sum of non-interacting single particles in momentum states. In contrast, interacting theories involving higher powers of fields can not be diagonalized, and for this reason their canonical quantization is believed to be impossible \cite{qft}. Our proposed procedure (\ref{modexp})-(\ref{en-omega}) diagonalizes the Hamiltonian of the strongly interacting field but, notably, this interaction is represented by the bilinear form $L_d=\frac{1}{\tau}\left(\phi_1\frac{\partial\phi_2}{\partial t}-\phi_2\frac{\partial\phi_1}{\partial t}\right)$ in (\ref{l1}) rather by the higher-order terms. The idea behind introducing parameter $\tau$ in $L_d$ is the same as in liquids. Although not derived from first-principles, the introduction of $\tau$ in liquid theory achieves two important results: (a) it simplifies an exponentially complex problem of coupled non-linear oscillators describing the motion of liquid particles in the anharmonic multi-well potential and thus provides a way to treat strongly-anharmonic interactions non-perturbatively \cite{ropp}; and (b) it quantifies an important and independently measurable liquid property (liquid relaxation time) directly linked to viscosity. This, in turn, enables developing a theory of liquid thermodynamics and provide relationships between different liquid properties \cite{ropp}. Similarly, introducing the dissipative term $L_d$ in (\ref{l1}) represents a way to treat strongly anharmonic self-interaction of the field non-perturbatively. Indeed, if the field self-interaction has a multi-well form in Fig. \ref{poten}, the field, in addition to oscillating in a single well, can move from one minimum to another (by either thermal activation or tunneling as discussed in, e.g., Ref. \cite{coleman}). This motion is completely analogous to diffusive particle jumps in the liquid responsible for the viscous term $\propto\frac{1}{\tau}$ in (\ref{gener3}) and the dissipative term $\propto\frac{1}{\tau}$ in (\ref{l1}). Therefore, $L_d$ describes the hopping motion of the field between different wells with frequency $\frac{1}{\tau}$. Note that this effect implies strong self-interaction of the field and can not be treated by usual perturbation methods in quantum field theory.

The important point from the above discussion is that the GMS and $\tau$ are related to the multi-well potential in Fig. \ref{poten}. We note that Figure \ref{poten} is a general construction giving rise to three regimes of particle dynamics and three states of matter: solids, liquids and gases. Particles oscillate in a single minimum in solids, oscillate and diffusively move between different minima in liquids and ballistically move above the potential barrier in gases.  We propose that similar regimes of field dynamics may exist in quantum field theory. In condensed matter physics, the potential barrier $U$ is set by the interatomic potential \cite{frenkel}. For self-interacting fields, we similarly assume that field self-interaction gives the landscape characterized by a single $U$.

We now return to our main results (\ref{q3})-(\ref{en-omega}) and other properties of our model. (\ref{en-omega}) can be viewed as the appearance of a dressed mass $m_d$ due to field self-interaction:

\begin{equation}
m_d=\sqrt{m^2-\omega_{\rm F}^2}
\label{dress}
\end{equation}
\noindent where the difference between $m_d$ and $m$ can be large, as expected from our non-perturbative approach.

This interpretation holds as long as $\omega_{\rm F}<m$. When $\omega_{\rm F}=m$, the dissipation term annihilates the mass in the dispersion relation (\ref{en-omega}) which becomes photon-like and gapless. When $\omega_{\rm F}>m$, the gap in momentum space opens up, as in (\ref{omega}) (see Fig. 1), and increases with $\omega_{\rm F}$:

\begin{equation}
p_g=\sqrt{\omega_{\rm F}^2-m^2}
\label{pg}
\end{equation}

We now discuss an interplay between real and imaginary terms in (\ref{q3}). These correspond to the energy and decay (half-width) of particles \cite{sudarshan,sud1}. The important effect concerns the crossover between propagating and non-propagating modes (PNM). The PNM crossover can be approximately defined as the equality between the mode period and decay time. In terms of energy, this corresponds to equality of $E_p$ and $\omega_{\rm F}$ in (\ref{q3}) (for simplicity, we assume that the energy of anti-particles is small compared to that of particles). Three regimes of the dissipative field dynamics follow.

In the first regime, all modes in (\ref{modexp}) remain propagating despite dissipation. Indeed, the energy gap in (\ref{en-omega}), $E_g=E_p(p=0)=\sqrt{m^2-\omega_{\rm F}^2}$, is larger than $\omega_{\rm F}$ if $\omega_{\rm F}<\frac{m}{\sqrt{2}}$. Under this condition, $E_p>\omega_{\rm F}$ for any $p$, as illustrated in Figure 1.

In the second regime, $\omega_{\rm F}$ increases so that $\omega_{\rm F}>\frac{m}{\sqrt{2}}$ but remains small enough so that $\omega_{\rm F}<m$ and the energy gap still exists as discussed above, i.e. $\omega_{\rm F}<m<\omega_{\rm F}\sqrt{2}$. In this case, the energy gap is smaller than $\omega_{\rm F}$: $E_g<\omega_{\rm F}$, and the PNM crossover takes place for all modes with $E_p<\omega_{\rm F}$ (see Fig. 1). Accordingly, all modes with momenta with $0<p<\sqrt{2\omega_{\rm F}^2-m^2}$ become non-propagating and do not contribute to the system energy.

In the third regime $\omega_{\rm F}>m$, the gap in momentum space opens up, and propagating modes start with $p>p_g$ (see (\ref{pg})). In this case, the PNM crossover implies that all modes with momenta in the range $p_g<p<\sqrt{2\omega_{\rm F}^2-m^2}$ become non-propagating (see Fig. 1). Together with the $p>p_g$, this implies that the range of non-propagating modes is $0<p<\sqrt{2\omega_{\rm F}^2-m^2}$ as in the second regime.

In the second and third regimes, the infra-red divergences are removed from evaluating integrals over $p$ because an integration starts from a finite value.

Importantly, our model has an {\it ultraviolet} cutoff. In addition to $E_p$ and $\omega_{\rm F}$, there is a third energy scale in our model: the height of the potential energy barrier $U$ in Fig. 2. The formalism of creation and annihilation operators assumes the quadratic form of the potential which provides the restoring force for the oscillator. For the potential in Fig. 2, this applies as long as the energy is smaller than $U$. For the energy above $U$, the potential provides no restoring force, and the formalism no longer applies. Therefore, the upper integration limit in all quantities of interest is approximately $U$, removing ultraviolet divergences in calculations. This results from the form of the potential in Fig. \ref{poten} only (and not from other ingredients of the theory such as, e.g., quantization). Notably, the GMS and the ultraviolet cutoff are related: the gap in momentum space emerges due to a finite $\tau$ ($\omega_{\rm F}$) in (\ref{energydis}), (\ref{en-omega}) or (\ref{pg}) which, in turn, arises from the potential in Fig. \ref{poten} with a finite $U$.

In summary, we represented dissipation due to field self-interaction by a bilinear term as in liquids and derived a quantized theory with dissipation. We found that particle energies are reduced by the dissipation which promotes the gap in momentum space and that particle mass becomes dressed by dissipation. We discussed the interplay between the dissipative and mass terms, different regimes of field dynamics as well as ultraviolet and infrared cutoffs emerging.\\

I am grateful to S. Ramgoolam, M. Baggioli, A. Zaccone and S. Grozdanov for discussions and to the EPSRC for support.

\end{document}